\documentclass[12pt]{article}

\usepackage{pazha}
\tightenlines
\usepackage{lscape}
\usepackage{epsf}
\usepackage{flushrt}
\usepackage{graphicx}

\def\a{$^{\mbox{\small a}}$}
\def\b{$^{\mbox{\small b}}$}
\def\c{$^{\mbox{\small c}}$}
\def\d{$^{\mbox{\small d}}$}
\def\e{$^{\mbox{\small e}}$}

\begin{document}

{\it \small Astronomy Letters, Vol.30, No. 1, 2004, pp.50-57. 
Translated from Pis'ma v Astronomicheskii Zhurnal, Vol. 30, No. 1, 2004,
pp. 58-66. 
Original Russian Text Copyright \copyright 2004 by Lutovinov, Tsygankov,
Grebenev, Pavlinsky, Sunyaev.}

\rule [0.5mm] {16.5cm} {0.2mm}\\
\rule [5mm] {16.5cm} {0.2mm}\\ 

\bigskip

\sloppypar

\title{\bf Two Years of Observations of the X-ray Pulsar SMC X-1 with
  the~ART-P Telescope onboard the Granat Observatory}  
\author{A.A.Lutovinov\affilmark{1,2}, S.S.Tsygankov\affilmark{1},
  S.A.Grebenev\affilmark{1}, M.N.Pavlinsky\affilmark{1},
  R.A.Sunyaev\affilmark{1,2}}
 
\affil{{Space Research Institute, Moscow, Russia}$^1$\\
{Max-Plank-Institut f\"ur Astrophysik, Garching, Germany}$^2$\\
}

\abstract{We present the observations of the pulsar SMC X-1 with the ART-P
telescope onboard the Granat observatory. We investigate the variability of
the flux from the source on time scales of several tens 
of days. The intensity variation of the pulsar are shown to be consistent
with the presence of a periodicity in the system with a characteristic time
scale of $\sim61$ days. The precession of an inclined accretion disk, as 
indirectly confirmed by the absence of low-state pulsations, may be
responsible for the observed variability. The spectrum of the source is well
described by a power-law energy dependence of the photon flux 
density with a slope of $\sim1.5$ and an exponential cutoff at energies
above $\sim14-18$ keV. We estimated the inclinations between the planes of
the orbit and the accretion disk and the magnetic field of the neutron 
star.\\ 
\copyright 2004 MAIK "Nauka/Interperiodica".\\

}

\section*{INTRODUCTION}

The X-ray pulsar SMC X-1 located in one the
nearest galaxies, the Small Magellanic Cloud (SMC),
is one of the most intense and rapidly rotating accreting 
X-ray pulsars. The source, which was first
discovered in X-rays in 1971 (Price et al. 1971), is
a member of a high-mass binary together with the
companion star Sk160, a B0-type supergiant with
a mass of $17.2M_{\odot}$ and a radius of $18R_{\odot}$ (Reynolds
et al. 1993). On short time scales, its flux pulsates
with a period of about 0.71 s, while the neutron
star itself is eclipsed by its companion for $\sim15$ h
(the duration of the X-ray eclipse) with a period
$P_{orb}\simeq3.892$ days. Analyzing the results obtained
over the entire preceding history of observations of
this source, Levine et al. (1993) determined the
orbital parameters of the binary and established that
its orbital period is decreasing at a rate 
$\dot P_{orb}/P_{orb}=-3.35\times10^{-6}$~ $yr^{-1}$.
    Apart from the rotation period of the neutron star
and the orbital period, the observations are indicative
 of a 50-60 day cycle in the binary (Wojdowski
et al. 1998), during which the X-ray luminosity of the
source varies between $10^{37}$~ erg s$^{-1}$ and several units
of $10^{38}$~ erg s$^{-1}$. The latter value is close to or even
higher than the Eddington limit for a spherically accreting 
neutron star with a mass of $1.4M_{\odot}$. This fact
as well as the observed steady spinup of the neutron
star for the more than 30 years since the discovery of
the pulsar (see, e.g., Wojdowski et al. 1998; Bildsten
et al. 1997; Lutovinov et al. 1994; and references
therein) suggest that a disk accretion model is most
likely realized in the Sk160/SMCX-1 system; the
accretion disk itself can be formed either from stellar
wind or during mass transfer through the inner 
Lagrangian point.
     Analyzing the EXOSAT light curves for the pulsar
 SMC X-1, Angelini et al. (1991) found a sharp
(by a factor of $\sim3$), short-time increase in the flux
from this source with a duration of $\sim80$ s, which
they interpreted as a type-II X-ray burst. Recent
RXTE observations of the pulsar have revealed more
than a hundred such events with time scales of several
 tens of seconds (Moon et al. 2003), suggesting
that the radiation and accretion mechanisms in the
source SMC X-1 under study and the pulsar/burster
GRO J1744­28 are similar.
     In this paper, we analyze the archival data obtained
with the ART-P telescope onboard the Granat observatory 
to study the long-term flux variations of the
source as well as its temporal and spectral parameters.

\section*{OBSERVATIONS}

   The ART-P X-ray telescope onboard the Granat
international orbital astrophysical observatory consists
 of four coaxial, completely independent units,
each of which includes a position-sensitive detector
with a geometric area of 625 cm$^2$ and an URA-based
coded mask. The telescope can image a selected region
 of the sky by a coded-aperture technique in a
$3^o.4 \times 3^o.6$~ field of view (the full beam width) with a
nominal resolution of $\sim5$~ arcmin in the energy range
3-60 keV. It has an energy resolution of $\sim22\%$~ in
the calibration 5.9-keV iron line. The observations
were carried out in the photon-flux mode. In this
mode, for each photon, its coordinates on the detector,
 the energy (1024 channels), and the arrival time
(the accuracy of the photon arrival time is 3.9 ms,
and the dead time is 580 µs) were written to buffer
memory. This mode allows one to perform both timing
and

\begin{landscape}
\small{
\begin{table*}[t]
\noindent
\centering
{\bf Table 1. }{ART-P observations of the pulsar SMC X-1 in 
 1990-1992 ÇÇ.}\\
\centering
\vspace{1mm}
\small{
\begin{tabular}{c|c|c|c|c|c|c}
\hline\hline
 Data & Dura-, & Orbital & Pulsation period, & Pulse fraction, & Flux,\c &  Luminosity,\b  \\
      & tion,s & phase   & s\a               & \% \b           &  mCrab  & $10^{38}$ erg s$^{-1}$  \\

\hline
05.04.90 & 30750 & 0.33-0.46 &           --             &   $5.1$\e     & $5.36\pm0.47$& $0.24\pm0.02$\\
22.04.90 & 34490 & 0.65-0.77 & $0.70957531\pm0.00000006$& $40.5\pm1.9$ & $64.0\pm1.0$ & $2.90\pm0.04$\\
24.04.90 & 12750 & 0.19-0.23 & $0.70957326\pm0.00000012$& $41.1\pm2.3$ & $63.0\pm1.2$ & $2.86\pm0.06$\\
25.04.90 &  8300 & 0.42-0.45 & $0.70957451\pm0.00000054$& $37.1\pm2.9$ & $63.3\pm1.4$ & $2.87\pm0.06$\\
11.05.90 & 17040 & 0.52-0.59 & $0.70955929\pm0.00000027$& $36.1\pm3.3$ & $58.9\pm1.3$ & $2.67\pm0.06$\\
12.05.90 & 44230 & 0.75-0.91 & $0.70955668\pm0.00000015$& $38.4\pm5.8$ & $27.2\pm0.7$ & $1.23\pm0.03$\\
25.01.91 &  8240 & 0.08-0.11 & $0.70932134\pm0.00000057$& $27.8\pm5.5$ & $18.8\pm3.5$ & $0.85\pm0.16$\\
13.04.91\d& 12060 & 0.09-0.14 & $0.70925884\pm0.00000223$& $16.0\pm7.2$ & $12.8\pm2.4$& $0.58\pm0.11$\\
14.04.91\d& 28340 & 0.38-0.46 & $0.70926034\pm0.00000087$& $11.7\pm2.1$ & $33.9\pm2.0$& $1.53\pm0.09$\\
21.04.91\d&  6250 & 0.13-0.16 & $0.70925315\pm0.00000140$& $18.7\pm3.5$ & $47.1\pm4.7$& $2.13\pm0.21$\\
22.04.91\d& 34320 & 0.43-0.56 & $0.70925311\pm0.00000014$& $18.1\pm2.0$ & $62.2\pm2.2$& $2.82\pm0.10$\\
17.05.92 & 15110 & 0.86-0.92 &           --             & $1.9$\e       &  $12.9$\e    &  $0.58$\e    \\
18.05.92 & 19200 & 0.17-0.26 & $0.70899141\pm0.00000623$& $4.2$\e &  $11.6$\e    &  $0.52$\e    \\

\hline
\end{tabular}
\vspace{3mm}

\begin{tabular}{ll}
\a & After correction for the motion of the Solar-System barycenter and the orbital motion in the binary. \\
\b & In the energy range 6­20 keV.\\
\c & In the energy range 6­20 keV for the assumed distance of d = 50 kpc to the source.\\
\d & For technical reasons, the spectrum of the source cannot be restored.\\
\e & The $3 \sigma$ upper limit for the pulse fraction is given in mCrab.\\
\end{tabular}}
\end{table*}
}
\end{landscape}
\noindent
spectral analyses of the radiation from each X-
ray source within the field of view of the telescope.
A detailed description of the telescope was given by
Sunyaev et al. (1990).

    The Granat observatory observed the region of the
SMC that contained the pulsar SMC X-1 in series
once a year (Table 1). A total of three series of 
observations with a total exposure time of $\sim271 000$~ s
were carried out for the source, which allowed us to
study in detail the radiation from the pulsar as well
as its spectrum and variability on various time scales.
Preliminary results of our timing analysis of the ART-
P/Granat observations of the pulsar SMC X-1 were
presented previously (Lutovinov et al. 1994). The
typical pulse profile for the source in the energy ranges
3-6, 6-10, 10-20, and 20-30 keV (Fig. 1) exhibits
two symmetric peaks located in the phase ranges
0.1-0.4 and 0.5-1.0, with the width of the second
peak decreasing with increasing photon energy.

    It should be noted that during the first series of
observations (in the spring of 1990), we used the first
module of the telescope. The subsequent observations
were performed with the third module, which had
a lower sensitivity at soft energies; therefore, all of
the results presented here refer to the energy range
6-20 keV. In addition, the high-voltage level of the
detector of this module varied significantly over time,
and it was not always possible to carry out calibration
observations of the Crab Nebula. As a result, we
ran into considerable difficulties in constructing the
response matrix for some of the observing sessions,
making it difficult or, in several cases, even impossible
to carry out a spectral analysis of the X-ray radiation
recorded with this module.

\section*{LONG-TERM INTENSITY VARIATIONS OF
               THE SOURCE AND BURSTS}

     As was noted above, the long-term observations of
\mbox{SMC~X-1} showed that, apart from the proper rotation
of the neutron star and its orbital motion, the system
has another, nearly periodic component-intensity
variability of the source on a time scale of several
days, with the recorded X-ray flux decreasing by more
than an order of magnitude. A similar pattern is also
observed in the other two well-known X-ray pulsars
\mbox{Her X-1} and \mbox{LMC X-4}, where it is attributable to
periodic eclipses of the emitting regions of the neutron
 star by a precessing accretion disk (Lutovinov
et al. 2000; La Barbera et al. 2001; and references
therein). It is assumed that the same mechanism may
also be responsible for the observed intensity variations
 in the pulsar SMC X-1.

   To determine the precession period $P_{prec}$~ of the
accretion disk, we fitted the ART-P data by a sinusoidal
 signal with a trial period varying in the range
of 40-80 days. The deviations of the measured fluxes
from their predicted values were determined by the
least-squares method. The best value was obtained
for a period $P_{prec}\simeq 61$~ days. This period agrees with
the results by Wojdowski et al. (1998), who found
long-term variations in the X-ray flux from SMC X-1
with a period of 50-60 days when simultaneously analyzing 
the ASM/RXTE and BATSE/CGRO data.
The limited set of data (observing sessions) makes it
impossible to completely cover the entire presumed
period. Therefore, our period estimate is not an accurate 
and statistically significant measurement of
the possible precession period $P_{prec}$~ (the statistical
significance of the peak on the periodogram is $\sim1.5\sigma$).
It most likely gives circumstantial evidence for the
presence of a third type of periodicity in the system
and the possible mechanism of its formation (see
below). 

The analysis of the light curve for the source by
Clarkson et al. (2003) shows that the presumed precession
 period is not constant in itself, but varies
smoothly over an interval of 40-60 days with a characteristic
 time scale of $\sim1600$~ days.

              Figure 2 shows the light curve for SMC X-1 
constructed over two years of ART-P observations in
the energy range 6-20 keV. The dots with ($1\sigma$) error
bars indicate the measured fluxes from the source in
mCrab during individual observing sessions, and the
solid line represents their best sinusoidal fit with a
period of $\sim61$~ days. It is undoubtedly of considerable
interest to compare the phases of the light curves
obtained simultaneously with the ART-P telescope
and the BATSE observatory. However, the quality of
the latter in this period was too low to make such a
comparison (see Fig.1 from Wojdowski et al. 1998).
A comparison with more recent RXTE observations
of the pulsar has revealed no correlation with our
results, which is most likely attributable to the 
variability of the precession period (see above).

   When analyzing the light curves for the presence
of X-ray bursts, we found several events that were
similar to those recorded by the RXTE observatory
(Moon et al. 2003) and that could be interpreted as
weak type-II bursts. Note that the statistical 
significance of the recorded events is low, $3-5\sigma$. However,
their roughly triangular shapes and HWHM of $\simeq13$~ s
obtained by Gaussian fitting are similar to those 
observed previously. Figure 3 shows the profile of one of
such bursts.

         \section*{SPECTRAL ANALYSIS}

   The spectrum of the X-ray pulsar SMC X-1 is
typical of this class of objects and can be described
by a simple power law. In general, depending on the
specific source, this model is modified by an exponential
 cutoff at high energies, absorption at soft energies,
 and emission or absorption lines. In the case of
SMC X-1, we did not detect any significant features
in emission or absorption. In addition, as our study
shows, the interstellar absorption estimated by different
 authors to be in a wide range, $5.9 \times 10^{20}-3.2 \times
10^{22}$~ atoms cm$^{-2}$~ (Wojdowski et al. 1998; Moon
et al. 2003), plays no important role in our analysis
of the radiation from the X-ray source in the energy
range 6-20 keV. Thus, the ultimate formula that we
used in our spectral analysis of the radiation from
SMC~X-1 is

\begin{equation}\label{1}
I(E)=I_{10}\,\left(\frac{E}{10\ \mbox{keV}}\right)^{-\alpha}
\left\{\begin{array}{cl} 1, &
\mbox{if}\ E<E_{c};\\ \exp{[-(E-E_{c})/E_{f}\,]},
&\mbox{if}\ E\geq E_{c},\\ 
\end{array}\right.
\end{equation}

Here, E is the photon energy in keV, $I_{10}$~ is the 
normalization of the power-law component to 10 keV, 
is the photon spectral index, $E_c$~ is the cutoff energy,
and $E_f$~ is the characteristic e-folding energy in the
spectrum of the source (although this is a purely
empirical formula with a cutoff at $E_c$, it is widely
used to fit the spectra of X-ray pulsars; see White
et al. (1983)).

   Figure 4 shows the photon spectra for the pulsar
(in photons cm$^{-2}$~ s$^{-1}$~ keV$^{-1}$) measured during several
 observing sessions with different intensity levels.
The curves represent the best-fit model spectra for the
source based on a simple power law or formula (1);
the best-fit parameters are given Table 2. In choosing
between a simple power law and its modification by a
cutoff at high energies, we used the $\Delta\chi^2$-test. Based
on this test, we determined the probability that it was
no accident that the $\chi^2$~ value improved when passing
to a more complex model. The statistical significance
of this passage is higher than 95\% in most cases.

     If we assume that the low state is observed when
the source is hidden behind the outer edge of a warped
or inclined (to the orbital plane) accretion disk and
that we see the flux attenuated by absorption and
scattering in the cold material on the disk periphery
or scattered in the hot corona above the disk (as is the
case in the system Her X-1 (Lutovinov et al. 2000)),
then, to a first approximation, the spectra of the high
and low states may be considered to differ only by
normalization and additional absorption. Fitting the
low-state spectrum of the pulsar (April 5, 1990) by
formula (1) with the parameters fixed at their high-
state values does not give a positive answer to the
above assumption (an appreciable increase in $\chi^2$~).
However, it cannot be completely rejected either, because 
the data are statistically limited. In this case,
an important criterion could be an increase in absorption
 at low energies. However, we cannot be certain
about this, because the energy range is limited: the
measured absorption in the low state is estimated to
be $(31.5\pm12.0)\times 10^{22}$~ atoms cm$^{-2}$, while the upper
limit for the absorption is $20\times 10^{22}$~ atoms cm$^{-2}$~
 ($1\sigma$).

     To investigate the spectral evolution of the source
on a scale of one pulse period, we performed phase-
resolved spectroscopy of the pulsar radiation in the
high state for two observing sessions, April 24 and
May 11, 1990. To this end, the data obtained during
each observing session were folded with the pulsation
period and divided into 12 time bins (the arrival time
of each photon was first corrected for the barycenter
of the Solar System and the motion of the neutron
star in the binary). The total accumulation time of
the signal for each of the 12 spectra was 1137 and
1568 s for the sessions of of April 24 and May 11,
respectively. The subsequent analysis indicated that
the spectra obtained in this way are well described by
a simple power law in the energy range 6-20 keV;
the spectral slope is virtually independent of the pulse
phase. This conclusion is illustrated by Fig. 5, which
shows a plot of the photon spectral index against the
pulse phase for the observing session of April 24.

\section*{DISCUSSION}

\section*{\em High and Low States}

It follows from the observations of SMC X-1 by
different observatories that the recorded X-ray flux
from this pulsar is not constant but undergoes large,
nearly periodic variations on a time scale of 50-
60 days. By analogy with the well-known binaries
SS 433, Her X-1, and LMC X-4, which exhibit a
similar pattern, the precession of an inclined accretion
disk is considered to be one of the main causes of the
observed variability. The normal companion Sk160 in
the binary under consideration is a supergiant with
an intense stellar wind that provides the bulk of the
accreting material forming 
\begin{landscape}
\small{
\begin{table*}[t]
\noindent
\centering
{\bf Table 2. Best-fit parameters for the spectra based on different models}{\a}\\
\centering
\vspace{1mm}
\small{

\begin{tabular}{c|c|c|c|c|c|c|c|c}
\hline\hline
&\multicolumn{3}{c|}{}&\multicolumn{5}{c}{}\\
Date&\multicolumn{3}{c|}{PL}&\multicolumn{5}{c}{PL+HEC}\\\cline{2-9}
    &$I_{10}$\b,&$\alpha$&$\chi^{2}_{\,N}(N)$\,\c&$I_{10}$,&$\alpha$&
    $E_{c}$,&$E_{f}$,&$\chi^{2}_{\,N}(N)$\\
    &$10^{-4}$  &        &                       &$10^{-4}$  &        &
    keV     &keV     &                    \\
\hline
05.04.90&$3.6\pm0.5$&$1.07\pm0.23$&1.28(11)&$4.4\pm0.9$&$0.39\pm0.54$&$12.9\pm3.5$&$6.3\pm4.4$&1.16(9)\\
22.04.90&$47.7\pm0.4$&$1.75\pm0.03$&1.43(32)&$49.2\pm0.6$&$1.67\pm0.04$&$17.0\pm2.8$&$24.9\pm14.0$&1.09(30)\\
24.04.90&$45.9\pm0.6$&$1.63\pm0.04$&1.35(34)&$47.9\pm0.8$&$1.52\pm0.05$&$18.1\pm3.0$&$15.4\pm10.7$&1.09(32)\\
25.04.90&$48.8\pm0.8$&$1.68\pm0.05$&1.14(26)&$50.7\pm1.1$&$1.59\pm0.06$&$19.7\pm3.9$&$13.3\pm13.3$&1.02(24)\\
11.05.90&$44.0\pm0.7$&$1.74\pm0.05$&1.37(32)&$47.8\pm1.3$&$1.49\pm0.09$&$13.7\pm1.8$&$13.2\pm4.7$&0.73(30)\\
12.05.90&$20.8\pm0.4$&$1.76\pm0.06$&1.27(32)&$22.6\pm1.2$&$1.53\pm0.16$&$10.9\pm3.2$&$24.8\pm10.9$&1.15(30)\\
25.01.91&$10.5\pm3.2$&$0.29\pm0.50$&0.45(7) &     --     &  --         &
--   & -- & -- \\

\hline
\end{tabular}}
\vspace{3mm}

\raggedright
\begin{tabular}{ll}
 \a & PL--a power-law spectrum, HEC--a high-energy cutoff.\\
 \b & The flux in photons cm$^{-2}$~ s$^{-1}$~ keV$^{-1}$~ measured at 10 keV. \\
 \c & The $\chi^2$ value normalized to the number $N$ of degrees of freedom.\\

\end{tabular}
\end{table*}
}
\end{landscape}

\noindent
the accretion disk. The
mechanism of its formation is not yet completely 
understood, because, in general, supergiants are close
to filling their Roche lobes and material can be 
transferred to the compact object through the inner 
Lagrangian point. Both mechanisms may operate in the
system simultaneously. It should be noted, however,
that, according to Clarkson et al. (2003), even if the
stellar wind plays a major role in the formation of the
disk, it is a collimated wind, and, thus, the two cases
are characterized by similar patterns of mass transfer
to the accretion disk.

           Larwood (1998) showed that a relation exists 
between the precession period and the binary parameters
 for a precessing accretion disk:

\begin{equation}
\frac {P_{orb}}{P_{prec}} = (3/7) q (1+q)^{-1/2}(R_{o}/{a})^{3/2} cos \delta,
\end{equation}

where $q$~ is the mass ratio of the normal component
 and the compact object; $R_o$~ is the outer radius
 of the accretion disk, which may be expressed
in fractions $\beta$~ of the Roche lobe size; $a$~ is the 
separation between the binary components; and $\delta$~ is
the angle between the orbital and disk planes. It
follows from an analysis of the optical light curves
for Sk160/SMC X-1 (Howarth 1982; Khruzina and
Cherepashchuk 1987) that the size of the accretion
disk is  $\beta\simeq0.7-1.0$. For the mass ratio $q = 10.8$~ and
periods $P_{orb}/P_{prec}\simeq 0.064$, we can estimate the range
of possible inclinations, $\delta\sim25^o-58^o$. The minimum
angle between the orbital and disk planes, $cos\delta\simeq1$,
corresponds to an accretion-disk size of $\beta\sim0.65$,
which slightly exceeds the maximum disk radius,  
$\beta\sim0.61$, obtained by Paczynski (1977) for accretion only
through the inner Lagrangian point. This discrepancy
is even larger if we take into account the range of 
possible precession periods. Thus, mass transfer from the
normal companion only through the inner Lagrangian
point in the system Sk160/SMC X-1 is unlikely.

   As was noted above, no X-ray pulsations were
found during the low state of the source, and only
upper limits for the pulse fraction were obtained 
(Table 1). However, since these limits are rather large, we
cannot completely rule out the presence of pulsations
in the low state. Wojdowski et al. (1998) presented
ROSAT measurements of the pulsation period in the
low state. However, they stipulate that these 
measurements were made at the very beginning of the
low state, when the source may have not yet been
completely shielded by the disk. A similar situation
was probably observed by the ART-P telescope on
May 18, 1992, when the intensity of the source was
low, but X-ray pulsations were recorded with low
statistical significance ($\sim4\sigma$).

\section*{\em Magnetic Field}

Based on the observed parameters of SMC X-1 obtained over a long
observing period, we can try to estimate the magnetic field of the neutron 
star in the system by using the model of an accretion disk
suggested by Li and Wang (1996).

The history of measurements of the pulsation period
for SMC X-1 is indicative of a virtually uniform
spinup of the neutron star since its discovery.
The mean rate of change of the period during
our observations in the period 1990-1992 was $\dot P =
-(32.6\pm±0.8)\times10^{-5}$~ s~yr$^{-1}$. Subsequent ROSAT,
ASCA, and RXTE observations of the source yield a
similar value, $\dot P = -(32.0\pm0.2)\times 10^{-5}$~ s~yr$^{-1}$~ 
(Wojdowski et al. 1998). Thus, assuming that the 
observed variability of the source is not related to the
variations in the intrinsic radiation from the pulsar,
but is determined by extraneous effects, we estimated
its bolometric luminosity from the observed spectral
parameters as $L_x = 47 \times 10^{37}$~ erg s$^{-1}$.

   In a situation where the neutron star spins up
almost uniformly over a long time interval, the change
in its rotation period is related to the binary 
parameters by the equation

\begin{equation}
\dot P = - \frac {\dot M (GMr_{0})^{1/2}n(\omega_{s})p^{2}} {2\pi I},
\end{equation}

where $\dot M$~ is the accretion rate; $I$~ is the moment of
inertia of the star; $r_0$~ is the inner edge of the 
accretion disk, which in our model is assumed to be
equal to the Alfven radius $r_A$; 
$n(\omega_{s})=1+ \frac {20(1-1.94\omega_{s})} {31(1-\omega_{s})}$~
is the dimensionless angular momentum; and $\omega_s =
\Omega_s/\Omega_k(r_0)$~ is the speed parameter. The latter depends
on the angular velocities $\Omega_s$~ and $\Omega_k(r_0)$~ of the neutron
star and the material at the inner edge of the accretion
disk, respectively. We estimated the moment of inertia
of the neutron star from the relation $I = 0.4M_xR^2$~
and assumed that its radius is $R = 10^6$~ cm, the mass
is $M_x = 1.4M_{\odot}$.

The magnetic moment $\mu$~ of the neutron star estimated using formula (3)
shows that the best agreement with the experimental data is achieved for
$\mu\simeq (0.05-0.1) \times 10^{30}$~ G cm$^3$, which is equivalent to a
neutron-star surface magnetic field of $\sim(1-2) \times 10^{11}$~G. This
conclusion about the relative weakness of the magnetic field is consistent
with the observation of bursts from the source that can be classified as
type-II bursts (see Moon et al. 2003). Moon et al. (2003) also established
that these bursts have much in common with the bursts observed previously
from another super-Eddington pulsar, GRO J1744-28. They suggested separating
out the so-called group of pulsars/bursters with magnetic fields of
$\sim10^{11}$~ G. Another important argument for this hypothesis is the
absence of cyclotron features in the spectrum of the source in the energy
range 3-100 keV, which corresponds to magnetic fields of
$\ge5\times10^{11}$~ G. 

Note that the applicability of the model under consideration at large is
limited by the fact that the speed parameter reaches the critical value
$\omega_c$ at which the angular momentum is not transferred to the 
neutron star. Having reached this level, the pulsar must pass from spinup to
spindown, which is not observed for SMC X-1. The fact that µ approaches
$10^{30}$~ G cm$^3$~ means that the speed parameter falls within the range 
of critical values $\omega_s$.

\section*{CONCLUSIONS}

    The X-ray pulsar SMC X-1 was repeatedly observed 
in 1990-1992 with the ART-P telescope onboard
 the Granat observatory. Over this period, we
accumulated the data that allowed us to investigate
the variability of the source on various time scales and
its spectrum and to estimate the binary parameters.
   
 We showed that, apart from the periodicities associated
 with the proper rotation of the neutron star and
the orbital motion, the system has yet another, nearly
periodic component that may be associated with the
precession of the accretion disk. Its period $P_{prec}\sim 
61$~ days, as inferred from the ART-P data, agrees well
with the observations of other observatories.

    The spectrum of the source is typical of X-ray
pulsars and can be described by a simple power law
with a cutoff at high energies. Our analysis showed
that the spectral shape depends weakly on the orbital
phase and intensity of the source. Phase-resolved
spectroscopy of the radiation from the source in the
high state revealed no dependence of the spectral
slope on the pulse phase either.

   The above estimates of the magnetic field for the
neutron star show that its strength must be $\sim10^{11}$~ G
to be consistent with the observational data (the
spinup rate the star, X-ray bursts, and the absence
of cyclotron features in the spectrum).

\bigskip

   This work was supported by the Russian Foundation 
for Basic Research (project no. 02-02-17347),
the Ministry of Industry and Science (grant no.
NSh-2083.2003.2 from the President of Russia),
and the Nonstationary Phenomena in Astronomy
Program of the Russian Academy of Sciences. We
are grateful to M. Revnivtsev for their discussions and
valuable remarks. We wish to thank Flight Director
K.G. Sukhanov; the staffs of the Lavochkin Research
and Production Center, RNIIKP, and the Deep Space
Communications Center in Evpatoria; the Evpatoria 
team of the Space Research Institute (Russian
Academy of Sciences); the team of I.D. Tserenin;
and B.S. Novikov, S.V. Blagii, A.N. Bogomolov,
V.I. Evgenov, N.G. Khavenson, and A.V. D'yachkov
from the Space Research Institute who operated the
Granat Observatory, provided the scientific planning
of the mission, and performed a preliminary processing
 of telemetry data. We also wish to thank the team
of M.N. Pavlinsky (Space Research Institute) and the
staff of the former Research and Development Center
of the Space Research Institute in Bishkek who
designed and manufactured the ART-P telescope.

\section*{REFERENCES}

\parindent=0mm

L. Angelini, L. Stella, and N. White, Astrophys. J. 371, 332 (1991).

A. La Barbera, L. Burderi, T. Di Salvo, et al., Astrophys.J. 553, 375
(2001). 

L. Bildsten, D. Chakrabarty, J. Chiu, et al., Astrophys.J., Suppl. Ser. 113,
367 (1997). 

I. Howarth, Mon. Not. R. Astron. Soc. 198, 29 (1982).

T.S. Khruzhina and A.M. Cherepashchyuk, Astron.Zh. 64, 345 (1987)
[Sov. Astron. 31, 180 (1987)]. 

J. Larwood, Mon. Not. R. Astron. Soc. 299, L32 (1998).

A. Levine, S. Rappoport, J. Deeter, et al., Astrophys.J. 410, 328 (1993).

X.-D. Li and Z.-R. Wang, Astron. Astrophys. 307, L5 (1996).

A.A. Lutovinov, S.A. Grebenev, R.A. Sunyaev, and M.N. Pavlinsky, Pis'ma
Astron. Zh. 20, 631 (1994) [Astron. Lett. 20, 538 (1994)].

A.A. Lutovinov, S.A. Grebenev, M.N. Pavlinsky, and R.A. Sunyaev, Pis'ma
Astron. Zh. 26, 803 (2000) [Astron. Lett. 26, 765 (2000)].

D.-S. Moon, S. Eikenberry, and I. Wasserman, Astrophys.J. Lett. 582, L91
(2003). 

F. Nagase, Publ. Astron. Soc. Jpn. 41, 1 (1989).

R. Price, D. Groves, R. Rodrigues, et al., Astrophys.J. Lett. 168, L7 (1971).

A. Reynolds, R. Hilditch, W. Bell, and G. Hill, Astron. Astrophys. 261, 337
(1993). 

E. Schreier, R. Giacconi, H. Gursky, et al., Astrophys.J. Lett. 178, L71
(1972). 

R. A. Sunyaev, S. I. Babichenko, D. A. Goganov, et al., Adv. Space Res. 10
(2), 233 (1990). 

N. White, J. Swank, and S. Holt, Astrophys. J. 270, 711 (1983).

P. Wojdowski, G. W. Clark, A. M. Levine, et al., Astrophys. J. 502, 253 (1998).

\bigskip
Translated by A. Dambis

\begin{figure}[t]
\includegraphics[width=16cm,bb=100 279 450 690,clip]{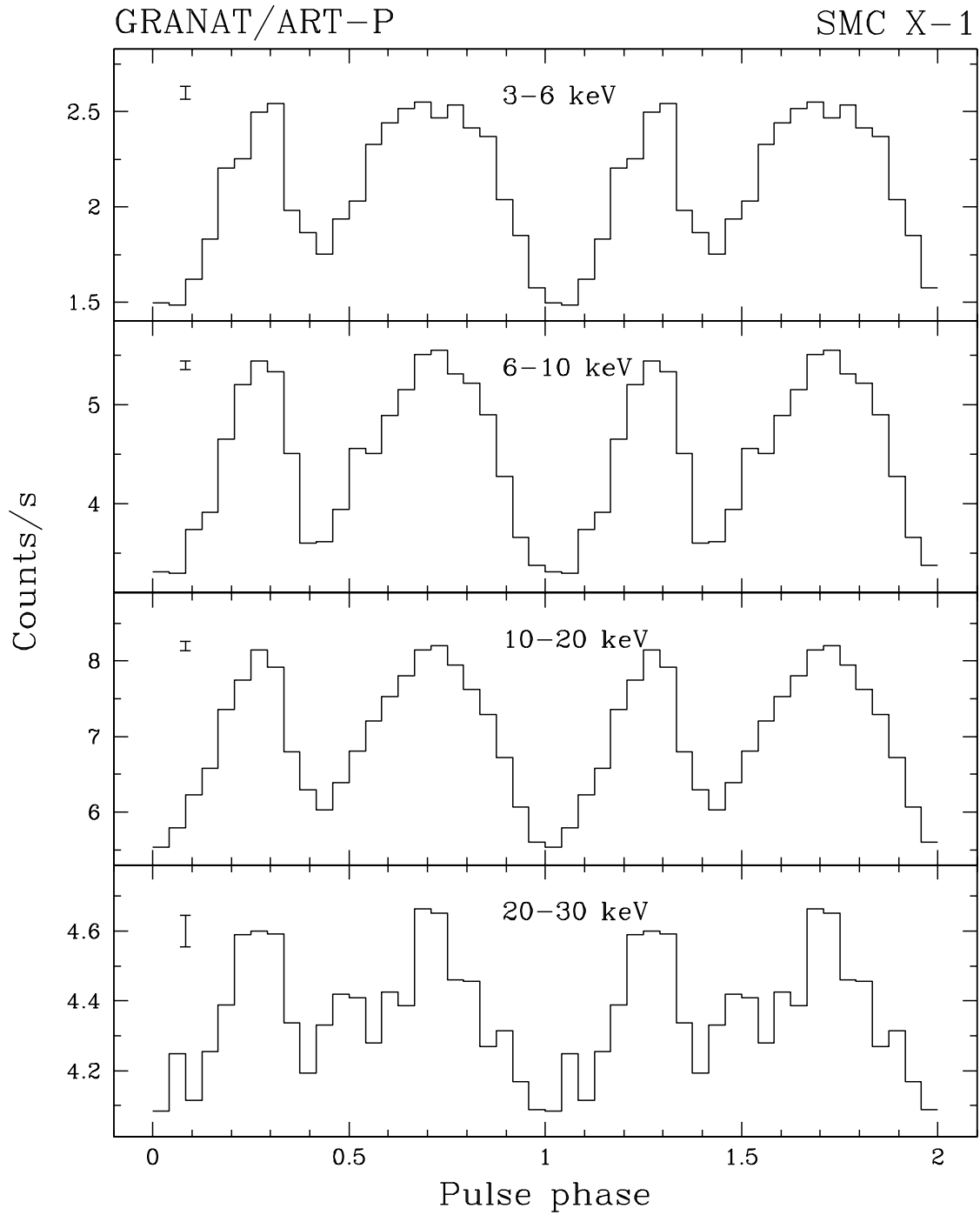}
\caption{Energy dependence of the pulse profile for
SMC X-1, as derived from the ART-P data of April 24,
1990. The errors correspond to one standard deviation.}
\end{figure}

\begin{figure}[t]
\includegraphics[width=16cm,bb=90 420 495 675,clip]{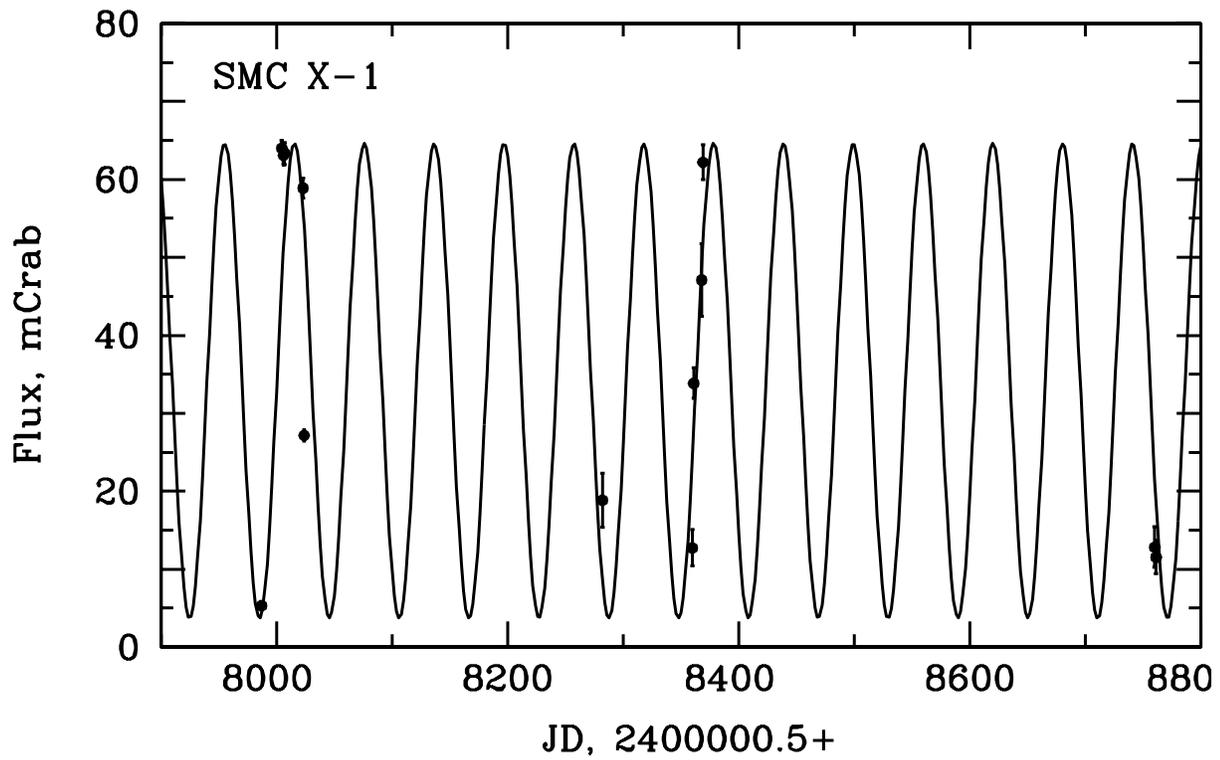}
\caption{The 6-20-keV light curve for SMC X-1 constructed 
over two years of ART-P observations. The dots
indicate the measured fluxes during an individual observation, 
and the solid line represents the best sinusoidal fit
with a 61-day period.}
\end{figure}

\begin{figure}[t]
\includegraphics[width=16cm,bb=100 418 500 690,clip]{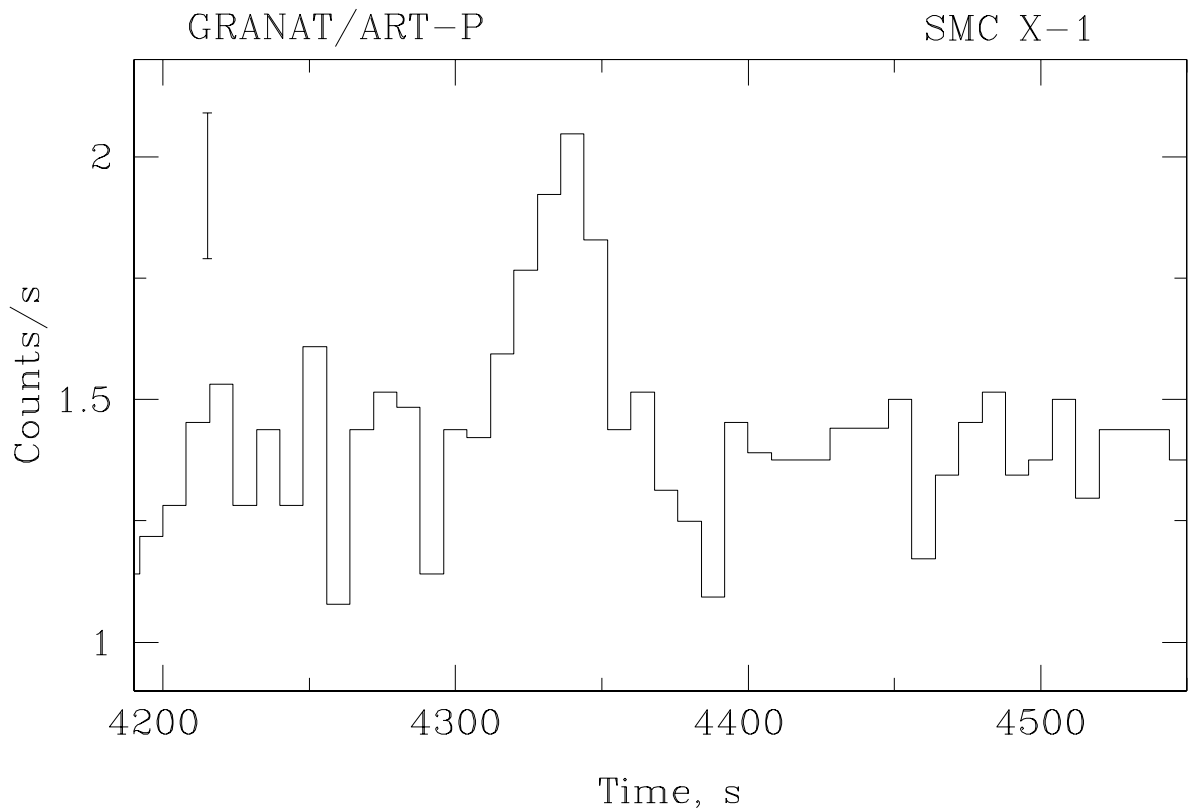}
\caption{The profile of a typical X-ray burst detected by
ART-P from SMC X-1 during the observing session of
April 22, 1990. Time in seconds from the beginning of the
observing session is along the horizontal axis. The time
bin is 8 s.}
\end{figure}

\begin{figure}[t]
\includegraphics[width=16cm,bb=100 370 430 700,clip]{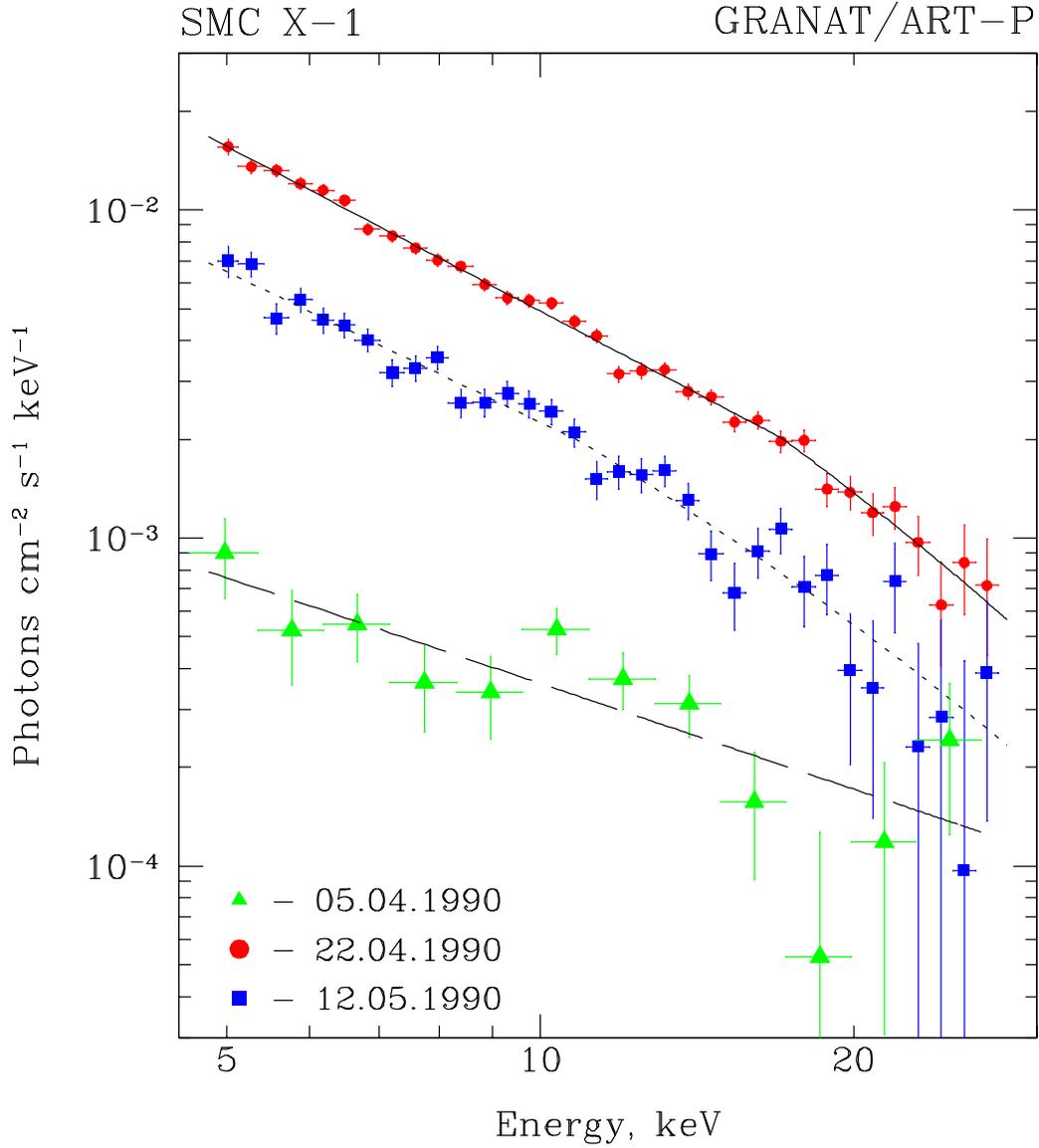}
\caption{The photon spectra for SMC X-1 measured with
the ART-P telescope during the first series of observations
 (the spring of 1990). Different symbols (circles,
squares, and triangles) indicate the photon spectra for the
source in states with different intensity levels; the curves
representtheir best fits by a power-law decrease in photon
flux density or formula (1).}
\end{figure}

\begin{figure}[t]
\includegraphics[width=16cm,bb=90 390 495 705,clip]{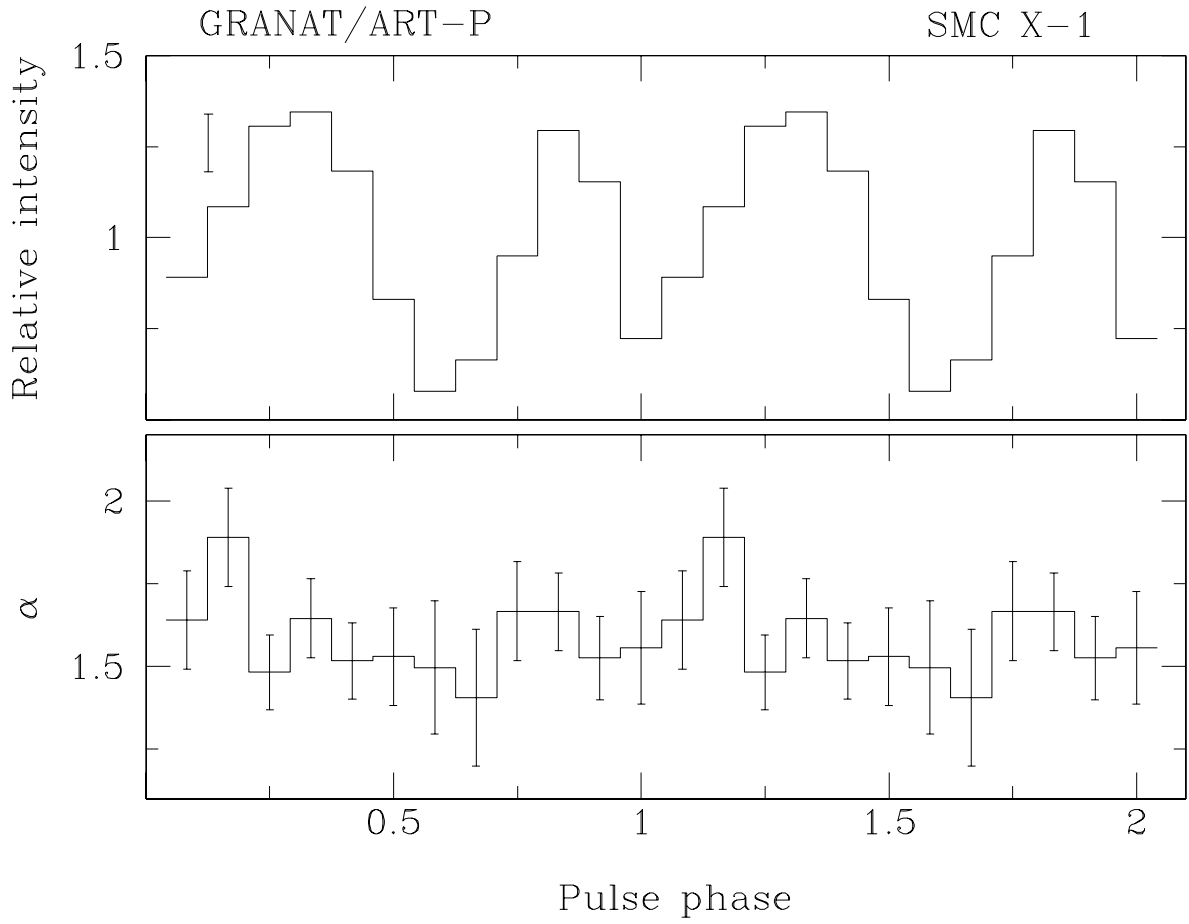}
\caption{Photon spectral index versus pulse phase. The
errors correspond to one standard deviation ($\sigma$).}
\end{figure}

\end{document}